\documentclass{acm_proc_article-sp-tweeked}

\usepackage[hyphens]{url}
\usepackage[boxed]{algorithm}
\usepackage{listings}
\usepackage{multirow}

\usepackage{color}
\definecolor{lightgray}{rgb}{.9,.9,.9}
\definecolor{darkgray}{rgb}{.4,.4,.4}
\definecolor{gray}{rgb}{.6, .6, .6}
\definecolor{purple}{rgb}{0.65, 0.12, 0.82}
\lstdefinelanguage{JavaScript}{
  keywords={break, case, catch, continue, debugger, default, delete, do, else, false, finally, for, function, if, in, instanceof, new, null, return, switch, this, throw, true, try, typeof, var, void, while, with},
  morecomment=[l]{//},
  morecomment=[s]{/*}{*/},
  morestring=[b]',
  morestring=[b]",
  ndkeywords={class, export, boolean, throw, implements, import, this},
  keywordstyle=\color{blue}\bfseries,
  ndkeywordstyle=\color{darkgray}\bfseries,
  identifierstyle=\color{black},
  commentstyle=\color{purple}\ttfamily,
  stringstyle=\color{red}\ttfamily,
  sensitive=true
}

\begin{document}

\title{Rules of Acquisition for Mementos and Their Content}

\numberofauthors{2}

\author{
%1st author
\alignauthor
Shawn M. Jones\\
	\affaddr{Los Alamos National Laboratory}\\
	\affaddr{Los Alamos, NM, USA}\\
	\email{smjones@lanl.gov}
% 2nd. author
\alignauthor
Harihar Shankar\\
	\affaddr{Los Alamos National Laboratory}\\
	\affaddr{Los Alamos, NM, USA}\\
	\email{harihar@lanl.gov}
}

\maketitle
\begin{abstract}
Text extraction from web pages has many applications, including web crawling optimization and document clustering.  Though much has been written about the acquisition of content from live web pages, content acquisition of archived web pages, known as mementos, remains a relatively new enterprise.  In the course of conducting a study with almost $700,000$ web pages, we encountered issues acquiring mementos and extracting text from them.  The acquisition of memento content via HTTP is expected to be a relatively painless exercise, but we have found cases to the contrary. We also find that the parsing of HTML, already known to be problematic, can be more complex when one attempts to extract the text of mementos across many web archives, due to issues involving different memento presentation behaviors, as well as the age of the HTML in their mementos.  For the benefit of others acquiring mementos across many web archives, we document those experiences here.

\end{abstract}

\category{H.3.7}{Digital Libraries}{Web Archives, Memento}[Systems Issues]
%\category{H.3.5}{Online Information Services}[Data sharing][Web-based Services]

\keywords{Digital Preservation, HTTP, Resource Versioning, Web Archiving, HTML}

\section{Introduction}

Text extraction of web page content is often done for web crawling optimization \cite{Manku:2007:DNW:1242572.1242592, Figuerola:2011aa} and document clustering \cite{6823554, Sandhya:2012aa}.  This paper documents the experience of comparing the text of archived web pages, hereafter referred to as \textbf{mementos}.  We believe that this is an unprecedented attempt at comparing mementos from almost $700,000$ web pages from across much of the lifespan of the web (1997-2012) and offers insights into the issues of using the content of different web archives for research purposes.  

Using the same data as \cite{10.1371/journal.pone.0115253}, we extracted more than 1 million URIs.  Because not all pages were archived sufficiently for our experiments (i.e. not enough mementos), we were left with 680,136 references.  Mementos from these remaining references come from more than 20 web archives.

Other studies in web page text extraction focused on comparing the text from downloaded live pages.  Studies exist that focus on extracting text from mementos in a single web archive \cite{AlNoamany2015, Jackson:2015}, but we are attempting to acquire and extract text from mementos across multiple web archives.  As a side-effect, we have discovered that each web archive modifies the content of mementos for the purposes of branding or user experience.  In terms of HTML, some mementos for the same live page are not presented consistently across different web archives.  These differences in the behavior of web archives with respect to the presentation of mementos and the challenges involved are the focus of this paper.

\section{Comparing Mementos}

When acquiring memento content, the following steps must be completed:
\begin{itemize}
\item \textbf{Acquisition of Web Resource}, where the URI of the web resource is dereferenced and the content then downloaded for use.  
\item \textbf{Text Extraction}, where the tags (for HTML) and control characters (for PDFs) are removed from the content, producing only text for review.  
\end{itemize}

We start with instances where downloading the content, given a URI, turns out to be more complex than anticipated.

\subsection{Acquisition of Web Resource}

Most mementos were easily acquired using cURL\footnote{\url{http://curl.haxx.se}} and standard HTTP against a given web archive.  Upon examination of the content of mementos acquired from some archives, however, it became clear that cases exist where an HTTP GET alone is not sufficient to acquire the content of a memento.  For WebCite, none of their mementos were directly accessible by using HTTP GET alone.  In addition, the Internet Archive replaces HTTP redirects with special pages that use JavaScript to simulate the redirect in a browser.  Also, we encountered pages where the redirection was embedded within the page content itself, noted here as \emph{HTML Meta Tag Refresh}.

\subsubsection{Acquiring Mementos From WebCite}

\begin{figure}
\centering
\caption{Example of a Failure In Dereferencing a Memento From WebCite}
\label{fig:web-citation-failure}
\includegraphics[width=0.5\textwidth]{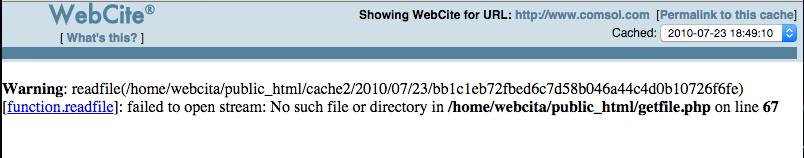}
\end{figure}

\begin{lstlisting}[language=JavaScript, float=*, breaklines=true, 
	caption={PhantomJS Script to Acquire Memento Content From WebCite},
	label={lst:phantomjs-script},
	showstringspaces=false,
	frame=single,
	rulecolor=\color{black}
	]

var page = require("webpage").create();
var system = require("system");

var wc_url = system.args[1];

page.onError = function(msg, trace) {
};

var response = [];

page.onResourceReceived = function(res) {
    response.push(res);
};

page.open(wc_url, function() {
    page.switchToFrame("main");
    console.log(page.frameContent);
    var headers = "";
    for (var i=0, r; r=response[i]; i++) {
        headers = "HTTP/1.1 " + r.status + "\n";
        if (r.url == "http://www.webcitation.org/mainframe.php") {
            for (var l=0, k; k=r.headers[l]; l++) {
                headers += k.name + ": " + k.value + "\n";
            }
            break;
        }
    }
    console.log("\n$$$$$$\n");
    console.log(headers);
    phantom.exit();
});

\end{lstlisting}

WebCite\footnote{\url{http://www.webcitation.org}} was created to address the link rot of web at large references and contains quite a few mementos for the URIs in our collection.  The first issue we encountered with WebCite was reliability.  As seen in the web browser screenshot of Figure \ref{fig:web-citation-failure}, visiting URIs on this site produces temporary errors.

Aside from these temporary errors, we observed that merely using cURL to dereference a WebCite URI does not result in the expected memento content.  Dereferencing two different WebCite URIs result in the exact same HTML content, as seen in Appendix \ref{adx:webcitation-examples}.  Based on our observations, the resulting HTML frameset loads the \texttt{topframe.php} and \texttt{mainframe.php} resources, then \texttt{mainframe.php} accepts a cookie containing session information to determine which content to load before redirecting the user to a page just containing the content in \texttt{mainframe.php}.  

We turned to PhantomJS\footnote{\url{http://phantomjs.org}} to help solve this problem; the script in Listing \ref{lst:phantomjs-script} handles this issue.  After developing this script, we then perceived rate limiting from WebCite.  We could find no documentation about rate limits, and requests for information from the WebCite webmaster were met with no response.  We overcame the rate limits by running several copies of the PhantomJS script from different networks and consolidating the results.  Due to the issues of using PhantomJS, which is slower than merely using cURL with HTTP GET, as well as the rate limiting, acquiring almost 100,000 mementos from WebCite took more than 1 month to complete.  For comparison, more than 1 million mementos were acquired from the Internet Archive in 2 weeks.

\subsubsection{Unresolved JavaScript Redirects}

\begin{figure}[t]
\centering
\caption{Example of a JavaScript Redirect from the Internet Archive}
\label{fig:ia-javascript-redirect-page}
\includegraphics[width=0.5\textwidth]{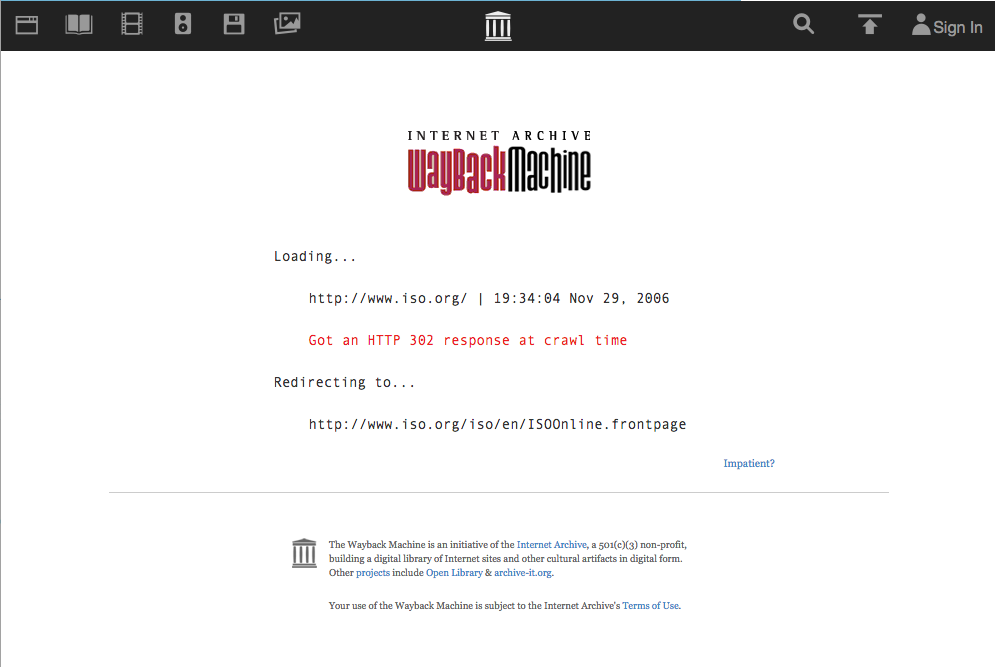}
\end{figure}

Most mementos from the Internet Archive fully support the expected HTTP status codes and the content from the majority of mementos was captured without incident.  Approximately 84,000 consisted of pages that resembled Figure \ref{fig:ia-javascript-redirect-page}.  Rather than merely issuing an HTTP 302 status code to redirect the web client, the Internet Archive has constructed these web pages that simulate a redirect in a web browser after some delay.

Once we detected these redirects, we used string matching to extract the redirect URIs from each page and then requested the content using that URI instead.  Unfortunately, some of these pages were a redirect to a redirect, and so content for almost 8,000 could not be acquired.  To solve this issue, one could create a PhantomJS client to download the target content of the chain of redirects.  Seeing as less than 1\% of mementos had additional JavaScript redirects, we excluded them from our dataset rather than creating this custom client.

\subsubsection{HTML Meta Tag Refresh} 

HTML includes the \texttt{meta}\footnote{\url{https://www.w3.org/TR/html-markup/meta}} element for representing metadata about a web page.  It provides the \texttt{http-equiv} attribute\footnote{\url{https://www.w3.org/TR/html-markup/meta.http-equiv.refresh.html#meta.http-equiv.refresh}}, allowing the web page author to either reload the current page after a certain number of seconds, or load a different page in place of the current page.  The use of the \texttt{http-equiv} attribute has been deprecated for this purpose\footnote{\url{https://www.w3.org/TR/WCAG10-HTML-TECHS/#meta-element}}, and is considered a bad practice for redirection because its use interferes with the history capability of some browsers\footnote{\url{https://www.w3.org/QA/Tips/reback}}.  Seeing as our mementos come from most of the history of the web, some exist from a time when this guidance was not given or well known.

In order to resolve this issue, one would need to load each page in a browser, or simulated browser, such as PhantomJS, and process the HTML on the page before being redirected.  Seeing as less than 1\% of mementos utilized this tag, we excluded them from our dataset rather than creating this custom client.  

\subsection{Text Extraction}

Acquiring memento content from web archives was difficult in some cases, but the resulting mementos contained a large variety of issues, some caused by the age of the pages and others by the web archives themselves.

We are also aware of the capability provided by Wayback installations, that allows access to the original archived content by inserting the \texttt{\_id} into a memento URI.  Not all archives use Wayback, so the issues encountered here still need to be addressed by anyone engaging in text extraction across archives.

To acquire the content from the mementos, we attempted to use the Python screen-scraping library BeautifulSoup\footnote{\url{http://www.crummy.com/software/BeautifulSoup/}}, but found it was unsuccessful in extracting text from some pages, so we instead relied upon the HTML module of the stricter lxml\footnote{\url{http://lxml.de/lxmlhtml.html}} library.  For PDF documents, we used the Python library PyPDF\footnote{\url{http://pybrary.net/pyPdf/}}.

Extracting only text was chosen for the following reasons:
\begin{enumerate}
\item We were aware that some web archives include additional HTML tags for branding and usability.  If two mementos come from different archives, then these additional tags and content could alter the results if compared.
\item When comparing mementos, not all similarity measures function the same way.  Some function at the byte level, while others rely upon more semantic constructs, like words and n-grams.  Some do not process tags, as seen in HTML, or control characters as found in PDFs.
\item One could extract the contents of PDFs and HTML documents for comparison or topic analysis.
\end{enumerate}

\subsubsection{Replacing BeautifulSoup By lxml}

We were aware that web archives included additional JavaScript and CSS in memento content for branding and usability purposes.  We originally attempted to use the Beautiful Soup code shown in Listing \ref{lst:bs4-script} to remove these items, but found that it did not work in all cases, often resulting in output containing no content.  An issue report was submitted to the Beautiful Soup maintainers in response to this issue\footnote{\url{https://bugs.launchpad.net/beautifulsoup/+bug/1489208}}.  Instead, we used the HTML module of the more strict lxml library, shown in Listing \ref{lst:lxml-script}, which was tested by visually inspecting the output of approximately 200 mementos.

\subsubsection{Removing Archive-Specific Additions}

\begin{lstlisting}[language=Python, float=t, breaklines=true, 
	caption={BeautifulSoup Code For Removing JavaScript and Stylesheets From Web Pages That Does Not Work In All Cases},
	label={lst:bs4-script},
	showstringspaces=false,
  	keywordstyle=\color{blue}\bfseries,
	ndkeywordstyle=\color{darkgray}\bfseries,
  	identifierstyle=\color{black},
  	commentstyle=\color{purple}\ttfamily,
  	stringstyle=\color{red}\ttfamily,
	frame=single,
	rulecolor=\color{black}
	]
soup = BeautifulSoup(content, "lxml")

# rip out JavaScript and CSS
for s in soup(["script", "style"]):
    s.extract()
    
# extract text
textonly = soup.get_text()
\end{lstlisting}

\begin{lstlisting}[language=Python, float=*, breaklines=true, 
	caption={lxml Code for Removing JavaScript and Stylesheets From Web Pages},
	label={lst:lxml-script},
	showstringspaces=false,
  	keywordstyle=\color{blue}\bfseries,
	ndkeywordstyle=\color{darkgray}\bfseries,
  	identifierstyle=\color{black},
  	commentstyle=\color{purple}\ttfamily,
  	stringstyle=\color{red}\ttfamily,
	frame=single,
	rulecolor=\color{black}
	]
fs = lxml.html.document_fromstring(content)
	
# rip out JavaScript
for element in fs.iter("script"):
    element.drop_tree()

# rip out CSS
for element in fs.iter("style"):
    element.drop_tree()

... removal of archive-specific tags here ...

# extract text
textonly = fs.text_content()
\end{lstlisting}

\begin{figure}[t]
\centering
\caption{Example of the Wayback Toolbar from \protect\url{http://web.archive.org/web/20081126132802/http://www.bnl.gov/bnlweb/pubaf/pr/PR_display.asp?prID=05-38}}
\label{fig:ia-wayback-toolbar}
\includegraphics[width=0.5\textwidth]{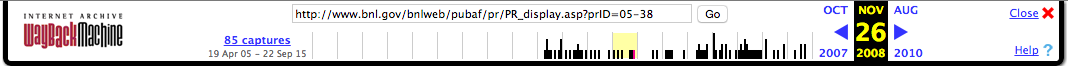}

\centering
\caption{Example of the National Archives Banner from \protect\url{http://webarchive.nationalarchives.gov.uk/20120405114247/http://www.decc.gov.uk/en/content/cms/statistics/publications/flow/flow.aspx}}
\label{fig:uk-national-archives-banner}
\includegraphics[width=0.5\textwidth]{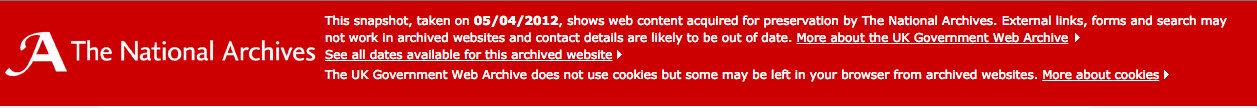}

\centering
\caption{Example of a memento from the PRONI archive, with archive-specific elements outlined in red, URI shown is \protect\url{http://webarchive.proni.gov.uk/20111214024729/http://eur-lex.europa.eu/LexUriServ/LexUriServ.do\%3Furi=CELEX:32008L0056:EN:NOT}}
\label{fig:proni-page}
\includegraphics[width=0.5\textwidth]{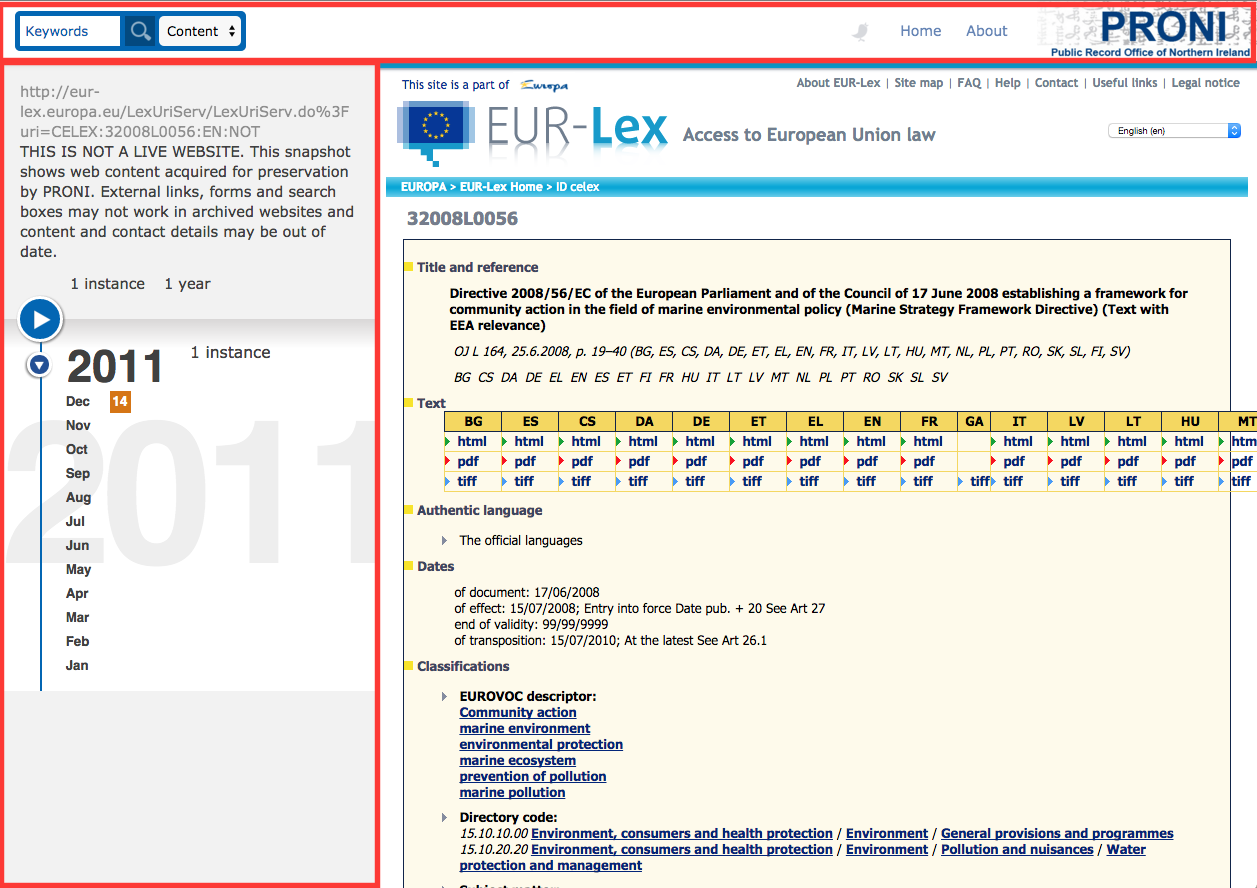}
\end{figure}

\begin{figure}[t]
\centering
\caption{Example of a memento from the Archive.is archive, URI shown is \protect\url{http://archive.is/19961226114737/http://www.rsinc.com/}}
\label{fig:archive-is-page}
\includegraphics[width=0.5\textwidth]{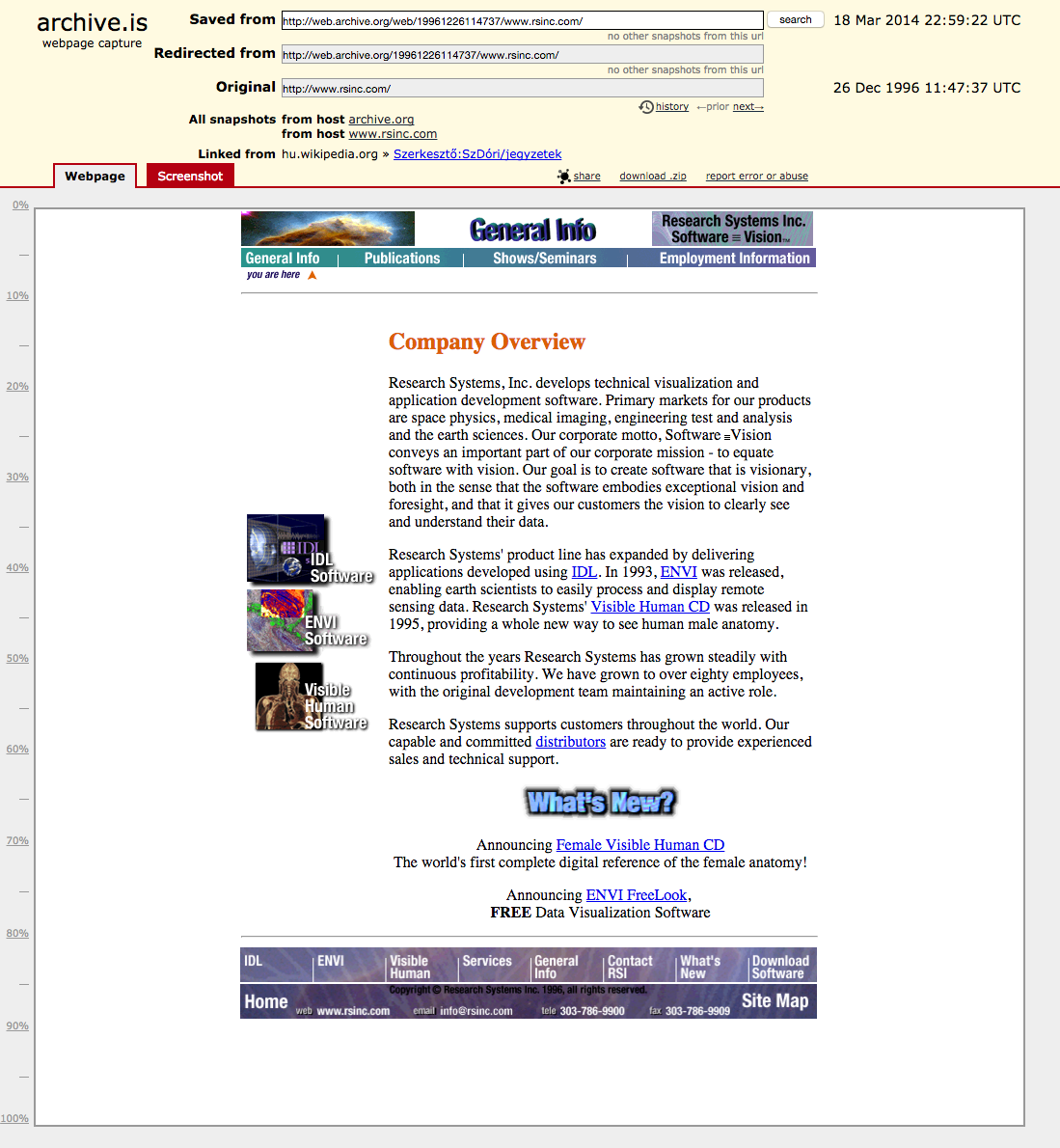}
\end{figure}

\begin{lstlisting}[language=Python, float=*, breaklines=true, 
	caption={Python Code for Removing Archive-Specific Elements From Web Pages},
	label={lst:archive-specific-elements},
	showstringspaces=false,
  	keywordstyle=\color{blue}\bfseries,
	ndkeywordstyle=\color{darkgray}\bfseries,
  	identifierstyle=\color{black},
  	commentstyle=\color{purple}\ttfamily,
  	stringstyle=\color{red}\ttfamily,
	frame=single,
	rulecolor=\color{black}
	]
# National Archives-specific stuff
for element in fs.iter("div"):
    if element.get("id") == "webArchiveInfobox":
        element.drop_tree()

# Wayback-specific stuff
for element in fs.iter("div"):
    if element.get("id") == "wm-ipp":
        element.drop_tree()

# PRONI-specific stuff
for element in fs.iter("div"):
    if element.get("id") == "PRONIBANNER":
        element.drop_tree()

# Archive.is
if '<meta property="og:site_name" content="archive.is"' in content:
    for element in fs.iter("div"):
        if element.get("id") == "HEADER":
            element.drop_tree()

    for element in fs.iter("table"):
        if element.get("id") == "hashtags":
            element.drop_tree()
            
textonly = fs.text_content()

# PRONI adds [ARCHIVED CONTENT]
# National Archives adds [ARCHIVED CONTENT]
if "PRONIBANNER" in content or "webarchiveInfobox" in content:
	if textonly[0:19] == "[ARCHIVED CONTENT] ":
		textonly = textonly[19:]

\end{lstlisting}

Once we had confirmed the more reliable behavior of the \texttt{lxml} library, we could then use it to remove items added by specific archives.  Using visual inspection and browser tools, we were able to identify specific blocks of HTML code that could be easily stripped from mementos without removing the actual content of the page.

Much of the archive-specific content can be removed by eliminating entire \texttt{div} elements, selected by their \texttt{id} identifier attribute, from the resulting memento.  

All Wayback\footnote{\url{http://www.netpreserve.org/openwayback}} sites (e.g. the Internet Archive\footnote{\url{http://archive.org/web/}}) add their banner, shown in Figure \ref{fig:ia-wayback-toolbar}, inside the division with the \texttt{wm-ipp} identifier.

The UK National Archives adds a banner, shown in Figure \ref{fig:uk-national-archives-banner}, inside the \texttt{div} identified by \texttt{webArchiveInfobox}. In addition, the string \texttt{[ARCHIVED CONTENT]}, appearing at the beginning after text extraction, must be stripped out of the resulting text.

The PRONI web archive adds a banner and a sidebar, outlined in red in Figure \ref{fig:proni-page}.  All of this added content is inside a division named \texttt{PRONIBANNER}.  The \texttt{[ARCHIVED CONTENT]} string also must be removed from the beginning of the extracted text. 

Archive.is, however is somewhat unique in its approach.  Figure \ref{fig:archive-is-page} shows a memento from the Archive.is archive, which places a header and sidebar on each memento, wrapping the memento content entirely and providing the reader with the ability to quickly scroll through a page by using the links on the right.  The bar and header are easily removed by eliminating the \texttt{div} elements with the identifiers \texttt{HEADER} and \texttt{hashtags}.  Seeing as the original web page content of the memento may use the \texttt{HEADER} identifier, the code shown in Listing \ref{lst:archive-specific-elements} only removes these elements if there is a corresponding meta tag identifying the memento as coming from Archive.is.

Appendix \ref{adx:html-archive-specific} contains example HTML snippets showing each of these sections for these archives.  In the case of other archives, they are either already covered by these cases, or the removal of all JavaScript and CSS eliminates any additional archive-specific content.

\subsubsection{Character Encoding Problems}

In order to properly compare the text of documents, they must be rendered properly so that characters can be consistently compared.  HTTP provides the \texttt{Content-Type} header \cite{rfc7231} to allow web clients to interpret the received content with the appropriate parsing and rendering software.  In order to compare documents, we relied upon the \texttt{Content-Type} header to indicate which library, lxml for \texttt{text/html} or pyPDF for \texttt{application/pdf}, is used to extract text from the given document.  For the content-type \texttt{text/html}, there is an additional \texttt{charset} attribute that helps a client render HTML in the correct character encoding.  Many character encodings exist\footnote{\url{http://www.iana.org/assignments/character-sets/character-sets.xhtml}}, allowing the browser to render different languages and other symbols.  For HTML documents, we additionally used the \texttt{charset} value to decode the content of a given resource so that it would be rendered correctly for comparison.  Decoding is necessary for document comparison and topic analysis.

Even though many character sets exist, UTF-8 \cite{rfc3629} is currently the most widely used\footnote{\url{https://googleblog.blogspot.com/2008/05/moving-to-unicode-51.html}}.  Even though the world has largely moved to UTF-8 for rendering all symbols, our dataset consists of web pages that exist before UTF-8 was adopted.  We observed that, for some of our mementos, web archives did not respond with the correct \texttt{charset} value, resulting in decoding errors.

To address this issue, we used the following algorithm to detect the character encoding\footnote{\url{https://en.wikipedia.org/wiki/Charset_detection}} for an HTML document.
\begin{enumerate}
\item Attempt to extract the character set has been specified by the \texttt{Content-Type} HTTP header
	\begin{enumerate}
	\item If a character set can be extracted, save it for use
	\item If no character set supplied, save character set as UTF-8 because UTF-8 is conceptually treated as a superset of several other character encodings\footnote{\url{http://www.unicode.org/standard/principles.html}}
	\end{enumerate}
\item Acquire the contents of the file using the saved character set
	\begin{enumerate}
	\item If the file is XHTML\footnote{\url{http://www.w3.org/TR/xhtml1/}}, attempt to extract the encoding from the XML encoding declaration and save it for use
	\item Try to decode the file contents using the saved character set
	\item If the saved character set fails, save character set as UTF-8 for the same superset reason
	\item If even UTF-8 does not work, throw an error
	\end{enumerate}
\end{enumerate}

Using this algorithm we were able to effectively evaluate most of the mementos.  Less than 500 still had decoding errors, but were discarded because they were less than 1\% of the whole.  It is possible that one could use confidence-based tools\footnote{\url{http://userguide.icu-project.org/conversion/detection}}, such as the chardet\footnote{\url{https://pypi.python.org/pypi/chardet}} library to guess the character encoding for these few that are left.

\subsubsection{Whitespace Inequality}

We learned that not all archives faithfully reproduce the archived content as it was originally captured.  In addition to adding additional tags and content for branding and usability, archives also do not necessarily preserve the spacing in an HTML document.  For example, Archive.is minifies\footnote{\url{https://en.wikipedia.org/wiki/Minification_(programming)}} all HTML, removing any additional space or newlines added by the author for legibility.  Because Archive.is removes this whitespace, the entire HTML document is presented as a single line, making direct content comparison with another archive impossible.  In contrast, the Internet Archive appears to reproduce the archived content very close to its original form, even including control characters added by HTML editors.

After text extraction, we replaced all whitespace with newlines for consistency in comparison between archives.

\begin{lstlisting}[language=html, float=*, breaklines=true, 
	caption={Simplified <noscript> Problem Example},
	label={lst:noscript-concept},
  	keywordstyle=\color{blue}\bfseries,
	ndkeywordstyle=\color{darkgray}\bfseries,
  	identifierstyle=\color{black},
  	commentstyle=\color{purple}\ttfamily,
  	stringstyle=\color{red}\ttfamily,
	frame=single,
	showstringspaces=false,
	rulecolor=\color{black},
	numbers=left,
	numberstyle=\tiny\color{gray},
	mathescape=true
	]

<html>
...
<body>
...

<noscript>  $\leftarrow$ lxml parser encounters this open tag
	<table border="0">
</noscript>

...  Much of the content for this page goes here ...

<noscript>
	</table>
</noscript> $\leftarrow$ lxml considers this to be the closing tag of the second noscript

</body>
</html> $\leftarrow$ lxml assumes that we will never get to the closing tag of the first noscript

\end{lstlisting}

\begin{lstlisting}[language=html, float=*, breaklines=true, 
	caption={Simplified <noscript> Example, With \&lt; and \&gt; as Seen in Mementos from Some Archives},
	label={lst:noscript-concept-ampsinstead},
  	keywordstyle=\color{blue}\bfseries,
	ndkeywordstyle=\color{darkgray}\bfseries,
  	identifierstyle=\color{black},
  	commentstyle=\color{purple}\ttfamily,
  	stringstyle=\color{red}\ttfamily,
	frame=single,
	showstringspaces=false,
	rulecolor=\color{black},
	numbers=left,
	numberstyle=\tiny\color{gray},
	mathescape=true
	]

<html>
...
<body>
...

<noscript>  $\leftarrow$ lxml parser encounters this open tag
	&lt;table border="0"&gt;
</noscript>

...  Much of the content for this page goes here ...

<noscript>
	&lt;/table&gt;
</noscript> $\leftarrow$ lxml considers this to be the closing tag because of the open table tag in between

</body>
</html>

\end{lstlisting}

\subsubsection{Null characters in HTML}

In attempting to extract text from mementos, we expected PDFs to contain control characters and noted that the pyPDF library processes them correctly.  We did not expect HTML to contain control characters, especially null characters.  In some cases, if these characters had not been removed, then individual tags could not be parsed by the lxml library.  In one example, a tag such as \texttt{<html>} was actually present in the file as \texttt{<[00]h[00]t[00]m[00]l[00]>}, where \texttt{[00]} represents a null character.  Because null characters are still characters, the parsing library did not detect a valid HTML tag if the null characters appeared inside one.

This issue was resolved by removing all null characters from HTML files prior to parsing them with lxml.

\subsubsection{Noscript With HTML-Encoded-HTML}

The \texttt{noscript} element\footnote{\url{https://www.w3.org/TR/html-markup/noscript.html}} is used to provide alternative content for user agents that do not support scripting.  Typically they present no issues to parsers, but in trying to process several mementos, we found that the lxml library could not extract text from $5,674$ references due to a problematic combination of HTML elements with the noscript tag.

HTML contains elements consisting of start tags, but no end tags.  The standard establishes a finite set of void elements\footnote{\url{https://www.w3.org/TR/html-markup/syntax.html#void-element}} (e.g. \texttt{img}, \texttt{hr}).  The \texttt{p} and \texttt{li} tags are also a special case whereby the end tag is optional\footnote{\url{https://www.w3.org/TR/html-markup/p.html}} \footnote{\url{https://www.w3.org/TR/html-markup/li}}.  Even the \texttt{body} end tag can be omitted in some cases\footnote{\url{https://www.w3.org/TR/html-markup/body}}.  Due to the optional nature of end tags for certain HTML cases, it appears that the lxml library accepts the \texttt{html} tag as the end of the document, even if some tags are not closed, and can arrive in a state where one can remove all content from a page by choosing the wrong element to remove.

This results in problems removing tags and extracting text, as shown in Listing 5.  Telling lxml to remove all \texttt{noscript} elements results in the removal of everything from line 7 to the end of the page because the parser ends up in a state where it cannot find the end of this first noscript tag.  One could just remove all tags in this example and extract the text, but if comparing the text with other mementos, we have found that some web archives do not present the contents of noscript elements the same way.  As seen in Listing \ref{lst:noscript-concept-ampsinstead}, some web archives present a memento for the same URI with the \texttt{<} and \texttt{>} replaced by \texttt{\&lt;} and \texttt{\&gt;}.  Because \texttt{\&lt;} and \texttt{\&gt;} are interpreted as text and not the opening and closing of a tag, text extraction returns these ``faux tags'', thus skewing any comparison one intends to do between mementos coming from archives that exhibit inconsistent presentation behavior.

Considering we encountered this scenario in less than 1\% of references, we discarded these items.  We did not find a parser that could handle this issue.

\section{Conclusions}

We were surprised to discover these issues during the acquisition and extraction of text from mementos.  Seeing as so many others have had success in comparing web pages, we assumed that mementos would be no different.  

Instead, we found that acquisition of mementos was problematic in approximately 25\% of our data set.  We dispelled the assumption that all web archives honor the HTTP protocol in full.  We observed that the Internet Archive, instead of preserving the original HTTP 302 status encountered when archiving a web page and creating a memento, generates a specialized HTML page with a JavaScript redirect that can only be followed once the content of the resulting specialized page is downloaded, processed, and executed.  We exposed that WebCite, an on-demand archive for preserving references, does not even return the memento content directly to the user, resulting in the need to again download, process, and execute content from a web page in order to acquire the content of a memento.  This behavior is not RESTful \cite{fielding2000architectural} because it requires content inspection and additional processing outside of HTTP to acquire a resource.  The requirement to utilize a web browser, rather than HTTP only, for the acquisition of web content is common for live web content, as detailed by Kelly \cite{Kelly2013} and Brunelle \cite{Brunelle2015}, but we did not anticipate that we would need a browser simulation tool, such as PhantomJS, to acquire memento content.

The acquisition of mementos is key to the success of many efforts, such as Memento \cite{rfc7089} and Hiberlink \cite{hiberlink_site}, that seek to combat link rot and content drift.  Most readers of web content are trying to access web archives for a better understanding of the context of the time period in which these resources were archived.  Problems accessing mementos by use of HTTP indicates the possibility of potential future issues with web browsers, preventing these readers from achieving their goal.

When processing the content of a memento, we exposed several issues.  Some mementos could not be processed by the popular HTML processing library Beautiful Soup, leading us to use lxml instead.  We documented how to remove archive-specific content for the archives in our experiment.  We found that character encoding can be inaccurate for mementos, resulting in the creation of a special algorithm for acquiring the character encoding, falling back to UTF-8 if that failed.  We observed that different web archives present mementos to the user with different levels of faithful reproduction, resulting in differences in whitespace.  We encountered mementos containing null characters that made parsing the HTML tags difficult.  Finally, we encountered mementos that could not be parsed by lxml at all because the DOM tree was corrupted by the use of \texttt{noscript} elements and unclosed tags.

It is obvious that some issues are a result of poor source material (e.g. the original page had improperly formatted HTML), but others, such as spacing, are introduced by the archives themselves.  Studies have been performed using text extraction on web pages, including mementos, but the effectiveness of text extraction is also quite important to the development of efforts for searching \cite{Brin:1998:ALH:297805.297827} and semantic analysis \cite{etzioni2005unsupervised} as well as archiving.  Without reliable support for text extraction, many of these efforts will be incomplete at best and fail at worse.

We have presented these issues to raise awareness so that future projects may benefit from these experiences and solutions.

\bibliographystyle{abbrv}
\bibliography{techreport}

\begin{thebibliography}{10}

\bibitem{AlNoamany2015}
Y.~AlNoamany, M.~Weigle, and M.~Nelson.
\newblock Detecting off-topic pages in web archives.
\newblock In S.~Kapidakis, C.~Mazurek, and M.~Werla, editors, {\em Research and
  Advanced Technology for Digital Libraries}, volume 9316 of {\em Lecture Notes
  in Computer Science}, pages 225--237. Springer International Publishing,
  2015.

\bibitem{Brin:1998:ALH:297805.297827}
S.~Brin and L.~Page.
\newblock The anatomy of a large-scale hypertextual web search engine.
\newblock In {\em Proceedings of the Seventh International Conference on World
  Wide Web 7}, WWW7, pages 107--117, Amsterdam, The Netherlands, The
  Netherlands, 1998. Elsevier Science Publishers B. V.

\bibitem{Brunelle2015}
J.~F. Brunelle, M.~Kelly, M.~C. Weigle, and M.~L. Nelson.
\newblock The impact of javascript on archivability.
\newblock {\em International Journal on Digital Libraries}, pages 1--23, 2015.

\bibitem{hiberlink_site}
P.~Brunhill, M.~Mewissen, T.~Strickland, R.~Wincewicz, C.~Grover, B.~Alex,
  R.~Tobin, C.~Metheson, and H.~{Van de Sompel}.
\newblock {hiberlink}.
\newblock \url{http://hiberlink.org}, July 2015.

\bibitem{etzioni2005unsupervised}
O.~Etzioni, M.~Cafarella, D.~Downey, A.-M. Popescu, T.~Shaked, S.~Soderland,
  D.~S. Weld, and A.~Yates.
\newblock Unsupervised named-entity extraction from the web: An experimental
  study.
\newblock {\em Artificial intelligence}, 165(1):91--134, 2005.

\bibitem{rfc3629}
{F. Yergeau}.
\newblock {UTF-8, a transformation format of ISO 10646, Internet RFC 3629}.
\newblock \url{http://tools.ietf.org/html/rfc3629}, November 2003.

\bibitem{rfc7231}
R.~Fielding and J.~Reschke.
\newblock {Hyptertext Transfer Protocol (HTTP/1.1): Semantics and Content,
  Internet RFC 7231}.
\newblock \url{http://tools.ietf.org/html/rfc7231}, June.

\bibitem{fielding2000architectural}
R.~T. Fielding.
\newblock {\em Architectural styles and the design of network-based software
  architectures}.
\newblock PhD thesis, University of California, Irvine, 2000.

\bibitem{Figuerola:2011aa}
C.~Figuerola, R.~D{\'\i}az, J.~Alonso~Berrocal, and A.~Zazo~Rodr{\'\i}guez.
\newblock Web document duplicate detection using fuzzy hashing.
\newblock In J.~Corchado, J.~P{\'e}rez, K.~Hallenborg, P.~Golinska, and
  R.~Corchuelo, editors, {\em Trends in Practical Applications of Agents and
  Multiagent Systems}, volume~90 of {\em Advances in Intelligent and Soft
  Computing}, pages 117--125. Springer Berlin Heidelberg, 2011.

\bibitem{Jackson:2015}
A.~N. Jackson.
\newblock {Ten years of the UK web archive: what have we saved?}
\newblock
  \url{http://anjackson.net/2015/04/27/what-have-we-saved-iipc-ga-2015/}, Apr
  2015.

\bibitem{6823554}
A.~Kavitha~Karun, P.~Mintu, and K.~Lubna.
\newblock Comparative analysis of similarity measures in document clustering.
\newblock In {\em Green Computing, Communication and Conservation of Energy
  (ICGCE), 2013 International Conference on}, pages 857--860, Dec 2013.

\bibitem{Kelly2013}
M.~Kelly, J.~F. Brunelle, M.~C. Weigle, and M.~L. Nelson.
\newblock {\em Research and Advanced Technology for Digital Libraries:
  International Conference on Theory and Practice of Digital Libraries, TPDL
  2013, Valletta, Malta, September 22-26, 2013. Proceedings}, chapter On the
  Change in Archivability of Websites Over Time, pages 35--47.
\newblock Springer Berlin Heidelberg, Berlin, Heidelberg, 2013.

\bibitem{10.1371/journal.pone.0115253}
M.~Klein, H.~Van~de Sompel, R.~Sanderson, H.~Shankar, L.~Balakireva, K.~Zhou,
  and R.~Tobin.
\newblock Scholarly context not found: One in five articles suffers from
  reference rot.
\newblock {\em PLoS ONE}, 9(12):e115253, 12 2014.

\bibitem{Manku:2007:DNW:1242572.1242592}
G.~S. Manku, A.~Jain, and A.~Das~Sarma.
\newblock Detecting near-duplicates for web crawling.
\newblock In {\em Proceedings of the 16th International Conference on World
  Wide Web}, WWW '07, pages 141--150, New York, NY, USA, 2007. ACM.

\bibitem{Sandhya:2012aa}
N.~Sandhya and A.~Govardhan.
\newblock Analysis of similarity measures with wordnet based text document
  clustering.
\newblock In S.~Satapathy, P.~Avadhani, and A.~Abraham, editors, {\em
  Proceedings of the International Conference on Information Systems Design and
  Intelligent Applications 2012 (INDIA 2012) held in Visakhapatnam, India,
  January 2012}, volume 132 of {\em Advances in Intelligent and Soft
  Computing}, pages 703--714. Springer Berlin Heidelberg, 2012.

\bibitem{rfc7089}
H.~Van~de Sompel, M.~L. Nelson, and R.~Sanderson.
\newblock {HTTP Framework for Time-Based Access to Resource States -- Memento,
  Internet RFC 7089}.
\newblock \url{https://tools.ietf.org/html/rfc7089}, December 2013.

\end{thebibliography}

\clearpage

\appendix
\clearpage

\section{WebCite Examples}
\label{adx:webcitation-examples}

\begin{lstlisting}[language=html, float=*htbp, breaklines=true, 
	caption={HTML From WebCite URI \protect\url{http://www.webcitation.org/6BToD7SUd}},
	label={lst:webcitation-frameset-1},
  	keywordstyle=\color{blue}\bfseries,
	ndkeywordstyle=\color{darkgray}\bfseries,
  	identifierstyle=\color{black},
  	commentstyle=\color{purple}\ttfamily,
  	stringstyle=\color{red}\ttfamily,
	showstringspaces=false,
	frame=single,
	rulecolor=\color{black}
	]

<!DOCTYPE html PUBLIC "-//W3C//DTD XHTML 1.0 Frameset//EN" "http://www.w3.org/TR/xhtml1/DTD/xhtml1-frameset.dtd">
<html xmlns="http://www.w3.org/1999/xhtml" xml:lang="en" lang="en">
  <head>
    <meta http-equiv="Content-Type" content="text/html; charset=utf-8"/>
    <title>WebCite query result</title>
    <link rel="stylesheet" type="text/css" href="/basic.css" />
		<link rel="stylesheet" type="text/css" href="/nicetitle.css" />
    <script src="https://www.google.com/recaptcha/api.js" async defer></script>
  </head>
  				<frameset rows="60,*" frameborder="0">
					<frame src="./topframe.php" name="nav" noresize="noresize" marginwidth="0" marginheight="0" scrolling="no" />
					<frame src="./mainframe.php" name="main" noresize="noresize" marginwidth="0" marginheight="0" />
				</frameset>

			</html>
			
\end{lstlisting}

\begin{lstlisting}[language=html, float=*, breaklines=true, 
	caption={HTML From WebCite URI \protect\url{http://www.webcitation.org/5rRjzl9dY}, a different URI, but with output identical to Listing \ref{lst:webcitation-frameset-1}},
	label={lst:webcitation-frameset-2},
  	keywordstyle=\color{blue}\bfseries,
	ndkeywordstyle=\color{darkgray}\bfseries,
  	identifierstyle=\color{black},
  	commentstyle=\color{purple}\ttfamily,
  	stringstyle=\color{red}\ttfamily,
	showstringspaces=false,
	frame=single,
	rulecolor=\color{black}
	]

<!DOCTYPE html PUBLIC "-//W3C//DTD XHTML 1.0 Frameset//EN" "http://www.w3.org/TR/xhtml1/DTD/xhtml1-frameset.dtd">
<html xmlns="http://www.w3.org/1999/xhtml" xml:lang="en" lang="en">
  <head>
    <meta http-equiv="Content-Type" content="text/html; charset=utf-8"/>
    <title>WebCite query result</title>
    <link rel="stylesheet" type="text/css" href="/basic.css" />
		<link rel="stylesheet" type="text/css" href="/nicetitle.css" />
    <script src="https://www.google.com/recaptcha/api.js" async defer></script>
  </head>
  				<frameset rows="60,*" frameborder="0">
					<frame src="./topframe.php" name="nav" noresize="noresize" marginwidth="0" marginheight="0" scrolling="no" />
					<frame src="./mainframe.php" name="main" noresize="noresize" marginwidth="0" marginheight="0" />
				</frameset>

			</html>
			
\end{lstlisting}

\clearpage
\section{HTML Snippet Examples of Archive-Specific Content}
\label{adx:html-archive-specific}

\begin{lstlisting}[language=html, float=*, breaklines=true, 
	caption={HTML Snippet from Internet Archive URI \protect\url{http://web.archive.org/web/20081126132802/http://www.bnl.gov/bnlweb/pubaf/pr/PR_display.asp?prID=05-38}, showing the beginning of the wm-ipp section},
	label={lst:ia-wm-ipp},
  	keywordstyle=\color{blue}\bfseries,
	ndkeywordstyle=\color{darkgray}\bfseries,
  	identifierstyle=\color{black},
  	commentstyle=\color{purple}\ttfamily,
  	stringstyle=\color{red}\ttfamily,
	frame=single,
	showstringspaces=false,
	rulecolor=\color{black}
	]
...
<div id="wm-ipp" lang="en" style="display:none;">

<div style="position:fixed;left:0;top:0;width:100%!important">
<div id="wm-ipp-inside">
   <table style="width:100%;"><tbody><tr>
   <td id="wm-logo">
       <a href="/web/" title="Wayback Machine home page"><img src="/static/images/toolbar/wayback-toolbar-logo.png" alt="Wayback Machine" width="110" height="39" border="0" /></a>
   </td>
   <td class="c">
       <table style="margin:0 auto;"><tbody><tr>
       <td class="u" colspan="2">
       <form target="_top" method="get" action="/web/form-submit.jsp" name="wmtb" id="wmtb"><input type="text" name="url" id="wmtbURL" value="http://www.bnl.gov/bnlweb/pubaf/pr/PR_display.asp?prID=05-38" style="width:400px;" onfocus="this.focus();this.select();" /><input type="hidden" name="type" value="replay" /><input type="hidden" name="date" value="20081126132802" /><input type="submit" value="Go" /><span id="wm_tb_options" style="display:block;"></span></form>
       </td>
       <td class="n" rowspan="2">
           <table><tbody>
           <!-- NEXT/PREV MONTH NAV AND MONTH INDICATOR -->
           <tr class="m">
            <td class="b" nowrap="nowrap">
...
\end{lstlisting}

\begin{lstlisting}[language=html, float=*, breaklines=true, 
	caption={HTML Snippet from UK National Archives URI \protect\url{http://webarchive.nationalarchives.gov.uk/20120405114247/http://www.decc.gov.uk/en/content/cms/statistics/publications/flow/flow.aspx}, showing the beginning of the webArchiveInfobox section},
	label={lst:national-archives-webarchiveinfobox},
  	keywordstyle=\color{blue}\bfseries,
	ndkeywordstyle=\color{darkgray}\bfseries,
  	identifierstyle=\color{black},
  	commentstyle=\color{purple}\ttfamily,
  	stringstyle=\color{red}\ttfamily,
	frame=single,
	showstringspaces=false,
	rulecolor=\color{black}
	]

...
 <div id="webArchiveInfobox" style="display: block; min-width: 130%; margin-left:-100px; height: 110px; position: absolute; top: -110px; left: -100px; width:1400px; padding: 0px; margin: 0px; background-color: #CC0000; opacity: 1; overflow: hidden;">
    <script type="text/javascript">
      BANNER_TNA_VERSION="13/5/2014";
      function enforceBanner() {
        thebody = document.getElementsByTagName("body")[0];
        if (thebody != null && thebody.style != null) {
          //inject style to override body with id
          thebody.style.backgroundPosition = "0px 73px";
          thebody.style.position = "relative";
          thebody.style.marginTop = "105px";
          if (thebody.style.width.lastIndexOf("px") == -1)
            thebody.style.width = "100%";
        }
      }
      enforceBanner();
    </script>

    <div id="webArchiveLogo" style="display: block; width: 317px; height: 70px; float: left; margin: 15px 0px 5px 0px; background: url(/media/img/TNA_logo_white_201006.gif) no-repeat 2px 20px;">&nbsp;</div>

    <!-- the text content -->
    <div style="display: block; margin-top: 18px; height: auto; width: 915px; margin-left: 325px; margin-right:0px; color: white !important; font-family: Verdana, Arial, Helvetica, sans-serif; font-size: 12px; line-height: 14px; text-align: left; color: white;">
...
\end{lstlisting}

\begin{lstlisting}[language=html, float=*, breaklines=true, 
	caption={HTML Snippet from PRONI Web Archive URI \protect\url{http://webarchive.proni.gov.uk/20111214024729/http://eur-lex.europa.eu/LexUriServ/LexUriServ.do\%3Furi=CELEX:32008L0056:EN:NOT}, showing the beginning of the PRONIBANNER section},
	label={lst:pronibanner},
  	keywordstyle=\color{blue}\bfseries,
	ndkeywordstyle=\color{darkgray}\bfseries,
  	identifierstyle=\color{black},
  	commentstyle=\color{purple}\ttfamily,
  	stringstyle=\color{red}\ttfamily,
	frame=single,
	showstringspaces=false,
	rulecolor=\color{black}
	]
...
<div id="PRONIBANNER" style="display: block; position: absolute; width: 100%; height: 96px; margin: 0; padding: 0; top: 0; left: 0; z-index: 10; border: none; color: #18417f; background-color: #ffffff; border-bottom: 4px solid #18417f;">
    <div id="PRONILOGO" style="display: block; width: 400px; height: 96px; float: right; margin: 0px; padding: 0; background-image: url('/media/img/pronilogo.jpg'); background-repeat: no-repeat; background-position: right bottom;"><span class="ACCESSIBLE" style="visibility: hidden;">Public Record Office of Northern Ireland</span>
    </div>
    <div id="PRONIBANNERCONTENT" style="display: block; height: 100%; padding: 0px 0px 0px 20px; font-family: Helvetica, Arial, sans-serif !important; font-size: 10.5pt !important; text-align: left !important; line-height: 1.4em !important; overflow: hidden;">
        <div style="display: block; padding: 15px 0 2px 0; margin: 0; color: #18417f !important; font-family: Helvetica, Arial, sans-serif !important; font-size: 10pt !important; line-height: 1.40em !important; font-weight: bold !important; background-color: transparent !important">THIS IS NOT A LIVE WEBSITE
...
\end{lstlisting}

\begin{lstlisting}[language=html, float=*, breaklines=true, 
	caption={HTML Snippet from Archive.is URI \protect\url{http://archive.is/19961226114737/http://www.rsinc.com/}, showing the beginning of the HEADER section},
	label={lst:archive-is-header},
  	keywordstyle=\color{blue}\bfseries,
	ndkeywordstyle=\color{darkgray}\bfseries,
  	identifierstyle=\color{black},
  	commentstyle=\color{purple}\ttfamily,
  	stringstyle=\color{red}\ttfamily,
	frame=single,
	showstringspaces=false,
	rulecolor=\color{black}
	]
...
<div id="HEADER" style="font-family:Verdana,Arial,Helvetica;background-color:#FFFAE1;border-bottom:2px #B40010 solid;min-width:1028px"><div style="padding-top:10px"></div><table style="width:1028px;font-size:10px" border="0" cellspacing="0" cellpadding="0"><tr><td style="width:150px;text-align:center;vertical-align:top" rowspan="5"><span style="white-space:nowrap;color:black;margin:0px;cursor:pointer" onclick="window.location='http://archive.is/'"><div style="font-size:24px">archive.is</div><div style="font-size:12px">webpage capture</div></span></td><td style="text-align:right;padding:3px 3px 0 3px;white-space:nowrap;vertical-align:top;font-size:14px;font-weight:bold">Saved from</td><td style="text-align:right;padding:3px 3px 0 3px;white-space:nowrap;vertical-align:top"><form style="text-align:left;margin:0" action="https://archive.is/search/" method="get"><table cellspacing="0" cellpadding="0" border="0"><tr><td style="width:500px"><input style="border:1px solid black;height:20px;margin:0 0 0 0;padding:0;width:500px" type="text" name="q" value="http://web.archive.org/web/19961226114737/www.rsinc.com/"/><input type="hidden" name="t" value="1395183562435"/><div style="text-align:right;font-size:10px"><span style="white-space:nowrap;padding:0;margin:0;color:gray">no other snapshots from this url</span></div></td><td style="vertical-align:top"><input style="width:60px;height:20px;padding:0;margin:0 0 0 3px" type="submit" tabindex="-1" value="search"/></td></tr></table></form></td><td style="text-align:right;padding:4px 5px 2px 5px;font-size:14px;white-space:nowrap;vertical-align:top" rowspan="2">18 Mar 2014 22:59:22 UTC</td></tr><tr><td style="text-align:right;padding:3px 3px 0 3px;white-space:nowrap;vertical-align:top;font-size:14px;font-weight:bold">Redirected from</td>
...
\end{lstlisting}

\begin{lstlisting}[language=html, float=*, breaklines=true, 
	caption={HTML Snippet from Archive.is URI \protect\url{http://archive.is/19961226114737/http://www.rsinc.com/}, showing the hashtags section allowing a user to quickly scroll through a memento},
	label={lst:archive-is-hashtags},
  	keywordstyle=\color{blue}\bfseries,
	ndkeywordstyle=\color{darkgray}\bfseries,
  	identifierstyle=\color{black},
  	commentstyle=\color{purple}\ttfamily,
  	stringstyle=\color{red}\ttfamily,
	frame=single,
	showstringspaces=false,
	rulecolor=\color{black}
	]
...
<table id="hashtags" style="text-align:right;font-family:Verdana,Arial,Helvetica;font-size:10px" border="0" height="100%"><tr><td id="0%" style="vertical-align:top"><a style="color:#999999" href="#0%">0%</a></td></tr><tr><td id="5%" style="vertical-align:top"><a style="color:#999999" href="#5%">?</a></td></tr><tr><td id="10%" style="vertical-align:top"><a style="color:#999999" href="#10%">10%</a></td></tr><tr><td id="15%" style="vertical-align:top"><a style="color:#999999" href="#15%">?</a></td></tr><tr><td id="20%" style="vertical-align:top"><a style="color:#999999" href="#20%">20%</a></td></tr><tr><td id="25%" style="vertical-align:top"><a style="color:#999999" href="#25%">?</a></td></tr><tr><td id="30%" style="vertical-align:top"><a style="color:#999999" href="#30%">30%</a></td></tr><tr><td id="35%" style="vertical-align:top"><a style="color:#999999" href="#35%">?</a></td></tr><tr><td id="40%" style="vertical-align:top"><a style="color:#999999" href="#40%">40%</a></td></tr><tr><td id="45%" style="vertical-align:top"><a style="color:#999999" href="#45%">?</a></td></tr><tr><td id="50%" style="vertical-align:top"><a style="color:#999999" href="#50%">50%</a></td></tr><tr><td id="55%" style="vertical-align:top"><a style="color:#999999" href="#55%">?</a></td></tr><tr><td id="60%" style="vertical-align:top"><a style="color:#999999" href="#60%">60%</a></td></tr><tr><td id="65%" style="vertical-align:top"><a style="color:#999999" href="#65%">?</a></td></tr><tr><td id="70%" style="vertical-align:top"><a style="color:#999999" href="#70%">70%</a></td></tr><tr><td id="75%" style="vertical-align:top"><a style="color:#999999" href="#75%">?</a></td></tr><tr><td id="80%" style="vertical-align:top"><a style="color:#999999" href="#80%">80%</a></td></tr><tr><td id="85%" style="vertical-align:top"><a style="color:#999999" href="#85%">?</a></td></tr><tr><td id="90%" style="vertical-align:top"><a style="color:#999999" href="#90%">90%</a></td></tr><tr><td id="95%" style="vertical-align:top"><a style="color:#999999" href="#95%">?</a></td></tr><tr><td id="100%" style="vertical-align:bottom;height:12px"><a style="color:#999999" href="#100%">100%</a></td></tr></table>
...
\end{lstlisting}

\end{document}